\documentclass[aps,twocolumn,showpacs,amsmath,amssymb]{revtex4}

\usepackage{graphicx}% Include figure files
\usepackage{bm}% bold math
\usepackage{amsmath}
\usepackage{epsfig}

\newcommand{\ud}{\mathrm{d}}
\newcommand{\bu}{\mathbf{u}}
\newcommand{\RaBen}{Rayleigh-B\'{e}nard }

\begin{document}

\title{Estimating the State of Large Spatiotemporally Chaotic Systems}

\author{Matthew Cornick}
\author{Brian Hunt}
\author{Edward Ott}
 \affiliation{University of Maryland, College Park, MD, 20742}
\author{Michael F. Schatz}
 \affiliation{Center for Nonlinear Science and School of Physics, Georgia Institute of Technology, Atlanta, GA, 30332}

\begin{abstract}
Data assimilation refers to the process of obtaining an estimate of a system's
state using a model for the system's time evolution and a time series of
measurements that are possibly noisy and incomplete. 
However, for practical reasons, the high dimensionality of large spatiotemporally chaotic systems prevents 
the use of classical data assimilation techniques.
Here, via numerical computations on the paradigmatic
example of large aspect ratio \RaBen convection, we demonstrate the applicability of a
recently developed data assimilation method designed to circumvent this difficulty.
In addition, we describe extensions of the algorithm for
estimating unknown system parameters.  Our results suggest the potential usefulness of
our data assimilation technique to a broad class of situations in which
there is spatiotemporally chaotic behavior.
\end{abstract}


\maketitle

Estimation of the state of an evolving dynamical system from measurements that are
possibly noisy and incomplete is a prerequisite for prediction and control in
many situations.  Furthermore, scientific investigations of dynamical processes
often require inference of an evolving system state from data.
The term `data assimilation' refers to
the case where a model for the time evolution of the system is available
and is used in conjunction with incoming measurements to
estimate the evolving system state.
The Kalman filter~\cite{KalmanIntro,Kalman}
optimally solves the data assimilation problem for systems with linear dynamics.  
The classical adaptation of the Kalman filter to nonlinear systems is called the 
extended Kalman filter~\cite{EKFbook}.  
However, it requires inversion of $N \times N$ matrices,
where $N$ is the number of model variables~\cite{variables}.
As a consequence, the application of such a technique to large, dynamically high-dimensional
spatiotemporally chaotic systems is infeasible because
no existing computers are large or fast enough to do the required matrix inversions.  
Despite these difficulties, recent developments from the field of numerical weather 
prediction~\cite{EKF,whit,tippetal,bishop,anderson01,hamill,hout}
suggest the possibility of achieving good accuracy (as in a Kalman filter), but in a way
that is computationally feasible for large systems.

The purpose of this Letter is to use numerical simulations to demonstrate the potential usefulness of 
a new, weather-inspired data assimilation method, the local ensemble transform Kalman filter (LETKF),
to a broad class of high dimensional spatiotemporally chaotic physical processes.
For specificity we employ a particular paradigmatic example: spatiotemporally chaotic \RaBen convection.
Flows such as spiral defect chaos~\cite{sdc,convection} in the \RaBen problem are, perhaps, the 
best studied experimental examples of spatiotemporal chaos; nevertheless, many general aspects of
spatiotemporal chaos remain poorly understood.
The LETKF is motivated by the observation that, in typical
examples of spatiotemporal chaos, spatial regions much smaller than the system size
are accurately described by many fewer degrees of freedom than the full system.
With this in mind, the LETKF employs many independent data assimilations in a large set of heavily
overlapping regions.  Because these regions are relatively small, the individual regional
computations are not prohibitive.  In addition, the regional data assimilation computations
are independent of each other and can thus be done in parallel.
Furthermore, by use of a simple example~\cite{LEKF1etal,LEKF2etal} it was indicated that, if the size
of the individual regions employed in LETKF is not too small (but still small compared to the
total system size), then state estimates with accuracies virtually the same as those for a
classical Kalman filter technique (thus presumably of near optimal accuracy) can be achieved.
For details of the LETKF algorithm we refer the interested reader to 
Refs.~\cite{LEKF1etal,LEKF2etal,LETKF}.

In \RaBen convection, a horizontal fluid layer of thickness $d$ is confined
between a heated lower plate and a cooled upper plate.  The onset of
fluid motion occurs when buoyancy overcomes viscous dissipation and
thermal diffusion as the temperature difference between the plates
$\Delta T$ is raised above a critical value $\Delta T_c$. 
\RaBen convection is typically modeled using the Boussinesq equations~\cite{boussinesq},
which are commonly nondimensionalized with
temperature scaled by $\Delta T$, length scaled by $d$, and time
scaled by the vertical diffusion time $d^2 / \kappa$, where
$\kappa$ is the thermal diffusivity.  This system of units is used throughout this Letter.
We numerically solve the Boussinesq equations~\cite{model} applying realistic boundary
conditions $\bu = 0$ and $T = 0$ (conducting) on the walls of the region
boundary ($x^2+y^2 \leqslant \Gamma^2$, $|z| \leqslant \frac{1}{2}$).
We denote the temperature deviation from
the conducting static solution (which is linear in $z$)
as $T$ and the fluid velocity as $\bu$.

The Boussinesq equations have two dimensionless parameters, the reduced Rayleigh number
$\epsilon$ and the Prandtl number $Pr$,
\begin{equation}\label{params}
 \epsilon \equiv \frac{R-R_c}{R_c} = \frac{\Delta T - \Delta T_c}{\Delta T_c} \quad ,
 \quad Pr \equiv \frac{\nu}{\kappa}.
\end{equation}
Here $R=g \alpha d^3 \Delta T / \nu \kappa$ is the Rayleigh number, 
$R_c$ is the critical Rayleigh number at convective onset, $g$ is gravitational
acceleration, $\alpha$ is the thermal expansion coefficient, and $\nu$ is
the kinematic viscosity.  Fluid convection arises when $\epsilon > 0$.
In addition, the radius
$\Gamma$ of the disk in units of the cell depth $d$, also referred to as the aspect ratio,
specifies the geometry.
We focused our studies on $\Gamma = 20$, $\epsilon = 1$, $Pr = 1$~\cite{res}.

In experiments, flows are visualized using the shadowgraph method~\cite{shadowgraph}, 
an indirect measurement of the fluid's spatial temperature variation.
In typical experiments, due to its difficulty, measurement of
the fluid velocity field is not performed.  The so-called mean flow, defined here as
$\bar{\bu}(x,y) \equiv \int \bu_{\bot}(x,y,z) \, \ud z$
(though see~\cite{meanflow} for a more complete description),
has been shown, through the use of simulations, to play a significant role in the 
dynamics~\cite{meanflow}.  Here $\bu_{\bot}$ denotes the horizontal component of
the fluid velocity $\bu=\bu_{\bot}+u_z \hat{\boldsymbol{z}}$.
Our goal is to determine the full fluid state $(T(x,y,z), \bu(x,y,z))$, from a time series of
shadowgraph measurements at a finite number of horizontal (pixel) locations,
and we view this as a test case investigation of the general
usefulness of our technique for laboratory experiments on spatiotemporal chaos.

We begin by considering a system state vector 
$\boldsymbol{\xi}$ with $N$ components, for which we have a dynamical model,
$\boldsymbol{\xi}_{j+1} = \boldsymbol{G}(\boldsymbol{\xi}_{j})$.
Our $\boldsymbol{G}(\cdot)$ is an integration of the Boussinesq equations~\cite{model} from
a time $t_j$ to $t_{j+1}$ where $t_j \equiv j \Delta t$ and $t_1, t_2, \ldots$ are the times at which
state estimates are constructed (also the times at which a measurement is taken).
The model state $\boldsymbol{\xi}$ consists of the
variables $T$ and $\bu$ defined on the grid points of the cylindrical mesh used by the model.

We map the temperature field to the shadowgraph light intensity $I(x,y)$ with a map $\boldsymbol{M}$
using a relation derived from geometric optics~\cite{shadowgraph,geometricoptics}
\begin{equation}\label{H}
I(x,y) = \frac{I_\circ(x,y)}{1 - a \nabla^{2}_{\bot} \bar{T}(x,y)} \equiv \boldsymbol{M}[T(x,y,z)]. 
\end{equation}
Here, $\nabla^{2}_{\bot} \equiv \partial^2 / \partial x^2 + \partial^2 / \partial y^2$ is 
the horizontal Laplacian, and the temperature field $\bar{T}(x,y) \equiv \int T(x,y,z) \ud z$ is
vertically averaged.
$I_\circ(x,y)$ is the incident light intensity 
and $a \equiv 2 z_1 |dn/dT|$, where $n$ is the index of refraction of
the fluid, $z_1$ is the optical path length from the midplane of the fluid layer to the
image plane (in units of $d$), and $dn/dT$ is evaluated at the average fluid temperature.
Note that for (\ref{H}) to be a good approximation to a more correct physical optics treatment~\cite{trainoff}
we require $\|a \nabla^{2}_{\bot} \bar{T}\| \ll 1$,
thus $\boldsymbol{M}[\cdot]$ is only weakly nonlinear in $T(x,y,z)$.
We assume that shadowgraph measurements 
are of the form $\boldsymbol{M}[T]_{mn}+\epsilon_{mn}$ where $\boldsymbol{M}[T]_{mn}$ denotes the value 
of $\boldsymbol{M}[T(x,y,z)]$ at the location of pixel $(m,n)$, and the quantities $\epsilon_{mn}$ denote
the errors in the measurement (noise), which we take to be zero mean iid Gaussian random variables with
variance $\sigma^2$.

Most data assimilation algorithms are iterative, cycling
between a predict and update step once every time interval $\Delta t$.
In the update step, current measurements are used to update
(or correct) the prediction.  The predict step then propagates the updated state,
via the model, to the next measurement time (\emph{i.e.}, it is a short term forecast). 
If the method works as intended, 
this process will closely synchronize the experiment and the model by coupling them
via the measurements.  The LETKF operates on this basic principle~\cite{LETKF,LEKF1etal,LEKF2etal}.

In order to assess how well the LETKF is performing, we will compare it to a more naive
approach that we call Direct Insertion (DI).  At the time $t_j$ of the shadowgraph measurement $I_j(x,y)$,
the DI method updates the predicted temperature field $T_j^{p}(x,y,z)$ by adding a correction 
$\delta T_j(x,y,z)$ that is the unique field that is
quadratic in $z$, matches the boundary conditions at $|z| = \frac{1}{2}$, and for which the updated field 
$T_j^u(x,y,z)=T_j^p(x,y,z)+\delta T_j(x,y,z)$ satisfies $\boldsymbol{M}[T_j^u(x,y,z)]=I_j(x,y)$.
This gives the update
\begin{equation}\label{DIinvert} \nonumber
\delta T_j(x,y,z) = (\bar{T}_j^u(x,y)-\bar{T}_j^p(x,y)) \left[ (3/2) - 6z^2 \right],
\end{equation}
where $\bar{T}_j^u(x,y)$ is found by solving a Poisson equation,
\begin{equation}\label{DIpoisson}
\nabla^{2} \bar{T}_j^u(x,y) = \frac{1}{a} \left[ 1-\frac{I_\circ(x_c,y_c)}{I_j(x_c,y_c)} \right],
\end{equation}
and $(x_c,y_c)$ is the location of the closest pixel to $(x,y)$ that is observed.
Note that with DI the velocity is not updated, ($\bu_{j}^{u}(x,y,z) = \bu_j^{p}(x,y,z)$);
rather, it develops via coupling with the temperature during the simulation step.
The predicted field (which has a proper $z$-dependence) is only slightly affected
since, if the method is working properly and measurements
are sufficiently frequent, the correction $\delta T_j(x,y,z)$ is small.
This method is the most successful data assimilation technique we have
tested that does not use an update based on the Kalman filter.
It is meant to represent what one might try without knowledge of the
techniques described in Ref.~\cite{LEKF1etal,LEKF2etal,LETKF}.

Now we describe so-called \emph{perfect model} tests in which a time series of states, generated from
a Boussinesq simulation of one particular initial condition,
serves as a proxy for the evolution of an experimental system whose state we wish to infer.
Simulated shadowgraph measurements
of this time series are generated every $\Delta t = 1/4$ using~(\ref{H}) with the parameters $a=0.08$, $I_{\circ}(x,y)=0.5$,
and adding noise of variance $\sigma^2$ as previously described.
Measurements are made sparse by removing shadowgraph pixels, leaving only
those that lie on \emph{observation locations}.
We introduce the measurement density $\rho \equiv s / (\pi \Gamma^2)$,
where $s$ is the number of observation locations.  
When $\rho$ is not small ($\rho > 5$) we randomly and uniformly distribute observation locations over the
disk; while at low density ($\rho \le 5$) the observation locations
are placed on a Cartesian grid covering the disk $x^2+y^2 \leqslant \Gamma^2$
(giving more repeatable results when using sparse measurements).
The observation locations are fixed for the entire data assimilation process.

We apply the LETKF and DI methods to our simulated shadowgraphs
to approximately reconstruct the original time series of true states.  
Here we document their performance as a function of
measurement noise $\sigma$ and measurement density $\rho$.
Simulated shadowgraphs are assimilated at times $t_j$, $j=1 \ldots J$.  During this process
the DI and LETKF converge on an estimate of the system state ($J$ chosen large enough to ensure convergence).
At time $t_J$ assimilation is turned off and the final updated state estimate is used as an
initial condition for a long term forecast.

Performance is quantified via the RMS relative error of the temperature and mean flow,
$E_{T}(t) = \left[ \langle |T - T^{t}|^2 \rangle/\langle |T^{t}|^2 \rangle \right]^{1/2}$ and
$E_{\bar{\bu}}(t) = \left[ \langle |\bar{\bu} - \bar{\bu}^{t}|^2 \rangle/\langle |\bar{\bu}^{t}|^2 \rangle \right]^{1/2}$.
Here, $T^{t}(x,y,z,t)$ and $\bar{\bu}^{t}(x,y,t)$ are the true temperature and mean flow from the proxy experiment,
and $\langle \cdot \rangle$ indicates a spatial average.  The fields $T(x,y,z,t)$ and $\bar{\bu}(x,y,t)$ are
the long term forecast fields from either DI or LETKF.
Note that, in contrast with the case of a real physical experiment, the perfect model set-up used here has the advantageous
property that the exact true fields $\bar{\bu}^t$ and $T^t$ are known and hence available for evaluating the actual error $E_T$ and
$E_{\bar{\bu}}$.

\begin{figure}[h]
\epsfig{file=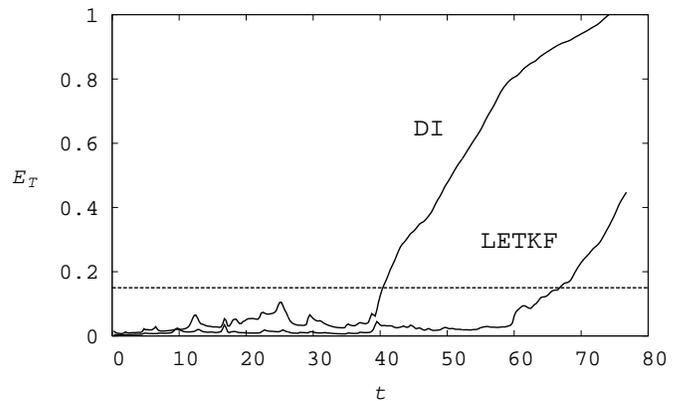,width=250pt}
\caption{ The error of the forecast temperature $E_T(t)$ with $\sigma=0.01$ and $\rho = 127$. 
	The dashed line is our chosen threshold defining the predictability time $E_T(t_p)=0.15$.}
\label{fig:idealTPRL}
\end{figure}

\begin{figure}[h]
\epsfig{file=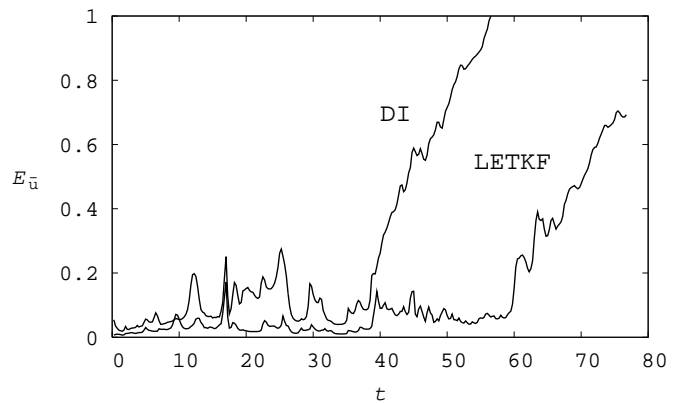,width=250pt}
\caption{ The error of the forecast mean flow $E_{\bar{\bu}}(t)$ (an unobserved variable) with $\sigma=0.01$ and $\rho = 127$.}
\label{fig:idealMFPRL}
\end{figure}

We define the \emph{ideal} scenario as measuring a shadowgraph with 
$\rho = 127$ (corresponding to a $451 \times 451$ shadowgraph image) and $\sigma = 0.01$
(this situation can be achieved in an experiment).
Under these conditions the DI and LETKF converge on a state estimate within
a few vertical diffusion times.
Both DI and the LETKF are effective for estimation of the (unobserved) mean flow $\bar{\bu}(x,y)$; however, the LETKF 
achieves an initial error $E_{\bar{\bu}}$ that is less than half that of DI.
The forecast error for a typical state estimate is shown in Figs.~\ref{fig:idealTPRL} and~\ref{fig:idealMFPRL}.
The general character of the forecasts is near-perfect agreement with the true state, followed by rapid divergence
due to local error growth at the location of a defect.
It is clear that the LETKF forecast is far superior to DI's, as measured by the 
\emph{predictability time} $t_p$,
defined as the time when the forecast error $E_T$ first crosses the (somewhat arbitrary) value of $0.15$.

\begin{figure}[h]
\epsfig{file=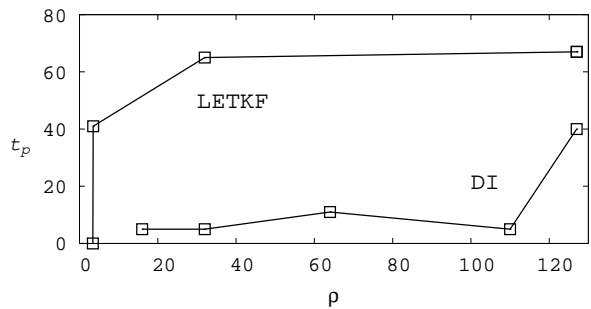,width=220pt}
\caption{ Comparison of DI and the LETKF; the predictability time $t_p$ is shown as the density of
	observations $\rho$ is reduced, demonstrating the superiority of the LETKF. }
\label{fig:perfectSPRL}
\end{figure}

Under non-ideal conditions the LETKF proves much more robust than DI.
Results for sparse measurements, shown in Fig.~\ref{fig:perfectSPRL}, demonstrate the large range of $\rho$ for which the LETKF converges.
One can observe the existence of a critical density of observations ($\rho \approx 1.3$)
below which it fails to converge.  DI on the other hand exhibits
a rapidly deteriorating forecast when even a few observation locations are removed.

\begin{figure}[!h]
\epsfig{file=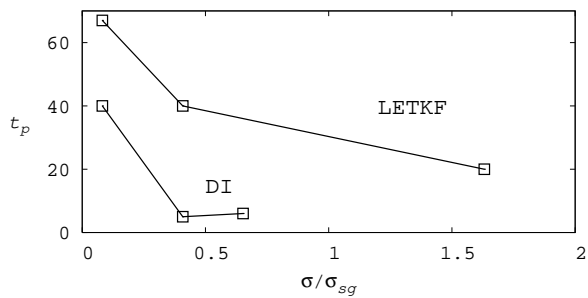,width=220pt}
\caption{ The predictability time $t_p$ is shown as measurement noise is increased.
	Noise levels shown are $\sigma=[0.01,0.05,0.08]$ for DI and $\sigma=[0.01,0.05,0.2]$ for the LETKF.}
\label{fig:perfectNPRL}
\end{figure}

We investigated performance as measurement noise was increased.
The meaningful \emph{signal to noise ratio} is $\sigma_{sg} / \sigma$,
where $\sigma_{sg} \equiv \langle I(x,y)- \langle I(x,y) \rangle \rangle$ is the
RMS intensity of a typical shadowgraph 
($\sigma_{sg} = 0.123$ when $a=0.08$ and $I_{\circ}(x,y)=0.5$).
DI relies on the Poisson solve~(\ref{DIpoisson}) which is fundamentally insensitive to noise
(it smoothes the right hand side).  However, this insensitivity competes with the sensitivity of the chaotic system dynamics
when producing forecasts.  The net result, in Fig.~\ref{fig:perfectNPRL} indicates that DI forecasts are only reliable for
a few vertical diffusion times when $\sigma > 0.4 \sigma_{sg}$,
whereas the LETKF yields accurate forecasts up to and exceeding $\sigma = \sigma_{sg}$.

Finally, we note that it is common for model parameters to be unknown, and that one of the
advantages of the Kalman filter methodology is that it can be utilized to infer unknown 
system parameters from measured time series~\cite{LETKF}.  In particular, let $\boldsymbol{p}$
denote the vector of unknown system parameters; with the model now determined by
$\boldsymbol{\xi}_{j+1} = \boldsymbol{G}(\boldsymbol{\xi}_{j},\boldsymbol{p})$.
One can then introduce an extended state space vector having the form 
$\boldsymbol{\gamma}=\left[\boldsymbol{\xi} \;\;  \boldsymbol{p}\right]^{\mathrm{T}}$,
where $\boldsymbol{p}$ is treated as a state variable with no time dependence.  
The extended model evolves as 
\begin{equation} \nonumber
\boldsymbol{\gamma}_{j+1}=
\left[ \begin{array}{cc}
	\boldsymbol{\xi}_{j+1} \\
	\boldsymbol{p}_{j+1} \end{array} \right] = \left[ \begin{array}{cc}
	\boldsymbol{G}(\boldsymbol{\xi}_{j},\boldsymbol{p}_{j}) \\
	\boldsymbol{p}_{j} \end{array} \right] = \hat{\boldsymbol{G}}(\boldsymbol{\gamma}_j).
\end{equation}
By now regarding the system model as $\hat{\boldsymbol{G}}$,
estimates of $\boldsymbol{\gamma}$ (and therefore the parameters $\boldsymbol{p}$)
result from an implementation in the same way
as for $\boldsymbol{\xi}$, but in the space of $\boldsymbol{\gamma}$ vectors.  
Using this method we can achieve estimates of
$\epsilon$ to within 0.02\% of the true value ($\epsilon = 1$) with the LETKF.
Remarkably, even when measurements are extremely sparse 
(\emph{e.g.}, $\rho = 3.6$, near the critical measurement density) the
$\epsilon$ estimates are within 0.2\%.  This demonstrates the utility of the
LETKF for inferring model parameters from incomplete data.

In conclusion, our results support the potential effectiveness of the LETKF at estimating the fluid state in 
laboratory \RaBen convection experiments.
We believe that the method we have presented is applicable to a large class of spatiotemporally
chaotic systems, and offers the possibility of bridging gaps between experiment and theory in the study of
spatiotemporal chaos~\cite{future}.

We are grateful to Laurette Tuckerman for the use of her Fortran code for simulating the Boussinesq equations.
This work was supported by the National Science Foundation (ATM 034225 and ATM 04-34193)
and the Office of Naval Research (Physics).

%\bibliography{RaBenLETKFPRL}

\begin{thebibliography}{24}
\bibitem[{var()}]{variables}
\bibinfo{note}{E.g., for a discretized partial differential model evolving $M$
  scalar spatial fields in time, the number of model variables is $M$
  multiplied by the number of grid points.}
\bibitem[{res()}]{res}
\bibinfo{note}{For these parameters we use a total of roughly $6 \times 10^5$
  grid points and a simulation time step of $0.01$.}
\bibitem[{fut()}]{future}
\bibinfo{note}{We note that in the work described here the model solves the
  same set of equations as the numerical proxy experiment (a Boussinesq
  simulation). In a real situation there is typically a discrepancy between the
  model evolution and that of the experimental system of interest. E.g., for
  the \RaBen case, our model would not be exact because of non-Boussinesq
  effects in an actual experimental fluid, as well as the finite time-stepping
  of the model, etc. Thus the results we have presented should be regarded as
  an indication of the best that could be expected. In a future publication we
  plan to test the technique using real (rather than simulated) shadowgraph
  measurements from a laboratory \RaBen experiment.}
\end{thebibliography}

\end{document}